\documentclass[10pt, journal]{IEEEtran}
\def\BibTeX{{\rm B\kern-.05em{\sc i\kern-.025em b}\kern-.08em
    T\kern-.1667em\lower.7ex\hbox{E}\kern-.125emX}}

\usepackage{amsmath,graphicx,psfrag,multirow,lscape}
\usepackage{amssymb, amsfonts, epsfig, latexsym, times}
\usepackage{graphics,cite,subfigure}
\usepackage{booktabs,balance}
\usepackage{comment}
\usepackage{subfigure}

\begin{document}
%
\title{Finite-State Markov Modeling of Leaky Waveguide Channels in Communication-based Train Control (CBTC) Systems}

\author{\IEEEauthorblockN{Hongwei Wang$^{\dag}$,~F. Richard Yu$^{\ddag}$, Li Zhu$^{\dag}$, Tao Tang$^{\dag}$, and Bin Ning$^{\dag}$}\\
$^{\dag}$State Key Laboratory of Rail Traffic Control and Safety, Beijing Jiaotong University, P.R. China\\
$^{\ddag}$Department of Systems and Computer Engineering, Carleton University, Ottawa, ON, Canada
}

\maketitle

\begin{abstract}
Leaky waveguide has been adopted in communication based train control (CBTC) systems, as it can significantly enhance railway network efficiency, safety and capacity. Since CBTC systems have high requirements for the train ground communications, modeling the leaky waveguide channels is very important to design the wireless networks and evaluate the performance of CBTC systems. In the letter, we develop a finite-state Markov channel (FSMC) model for leaky waveguide channels in CBTC systems based on real field channel measurements obtained from a business operating subway line. The proposed FSMC channel model takes train locations into account to have a more accurate channel model. The overall leaky waveguide is divided into intervals, and an FSMC model is applied in each interval. The accuracy of the proposed FSMC model is illustrated by the simulation results generated from the model and the real field measurement results.
\end{abstract}


\IEEEpeerreviewmaketitle


\section{Instruction}
Urban rail transit systems are developing rapidly around the world. Due to the huge urban traffic pressure, improving the efficiency of urban rail transit systems is in demand. A key sub-system of urban rail transit systems, communications-based train control (CBTC) is an automated train control system using train-ground communications to ensure the safe and efficient operation of rail vehicles \cite{What_is_communication-based_train_control}. It can enhance the level of  service offered to customers and improve the utilization of railway network infrastructure \cite{CBTC_Standard}.

As urban rail transit systems are built in a variety of environments (e.g., underground tunnels, viaducts, etc.), there are different wireless network configurations and propagation schemes. For the viaduct scenarios, leaky rectangular waveguide is a popular approach, as it can provide higher performance and stronger anti-interference ability than the free space \cite{Leaky_Waveguide_for_Train-to-Wayside_Communication-Based_Train_Control}, as the electromagnetic waves propagate inside the leaky waveguide in the longitudinal direction, which can eliminate the effects of the co-channel interference. In addition, due to the available commercial-off-the-shelf equipment, wireless local area networks (WLANs) are often adopted as the main method of train ground communications for CBTC systems \cite{ZHULI_FSMC,ZFBT12}.

Modeling the channels of urban rail transit systems is very important to design the wireless networks and evaluate the performance of CBTC systems. Although some excellent works have been done on modeling channels \cite{Characterization_of_Large_Scale_Fading_for_the_2_4_GHz_Channel_in_Obstacle_Dense_Indoor_Propagation_Topologies, 5_GHz_Band_Vehicle_to_Vehicle_Channels_Models_for_Multiple_Values_of_Channel_Bandwidth, Channel_Modeling_for_Vehicle_To_Vehicle_Communications}, few of them focus on the characteristics of leaky waveguide channels in CBTC systems.

In this paper, we develop a finite-state Markov channel (FSMC) model for leaky waveguide channels in CBTC systems, based on real field CBTC channel measurements obtained from the business operating Beijing Subway Yizhuang Line. Due to the good trade-off between accuracy and complexity, the FSMC model has been successfully employed in different channels, including Rayleigh fading channel \cite{Finite_state_Markov_channel_a_useful_model_for_radio_communication_channels}, Ricean fading channel \cite{Finite-state_Markov_modeling_of_correlated_Rician-fading_channels} and Nakagami fading channel \cite{Fast_simulation_of_diversity_Nakagami_fading_channels_using_finite-state_Markov_models}. The proposed FSMC channel model takes train locations into account to have a more accurate channel model. The accuracy of the proposed FSMC model is illustrated by the simulation results generated from the model and the real field measurement results. The effects of distance interval are also discussed.


%
%
%
%
%
%

\section{Overview of Communication-Based Train Control}
\label{OverviewCBTC}

Fig. 1 describes a CBTC system. In this system, continuous bidirectional wireless communications between each mobile station (MS) on the train and the wayside APs are adopted instead of the traditional fixed-block track circuit. The railway line is usually divided into areas or regions. Each area is under the control of a zone controller (ZC) and has its own radio transmission system. Each train transmits its identity, location, direction and speed to the ZC. The radio link between each train and the ZC should be continuous so that the ZC knows the locations of all the trains in its area all the time in order to guarantee train operation safety and efficiency.
Generally speaking, in viaducts scenarios, the performance of wireless communication could be affected by the interference caused by the other wireless devices from the surrounding buildings in the city,  and the consumer WiFi devices on the train can cause interference to CBTC systems as well. As a result, leaky waveguide has been adopted as the propagation medium in CBTC systems, such as Beijing Subway Yizhuang Line.  Considering the unique characteristics of leaky waveguide in CBTC systems, we propose an FSMC model of leaky waveguide.



\begin{figure}[tp]
	\centering
	\includegraphics[width=0.4\textwidth]{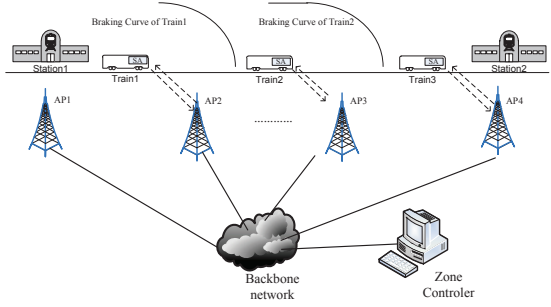}  \label{aCBTC}
	\caption{A communication-based train control (CBTC) system.}
\end{figure}

\section{Real Field Measurements of Leaky Waveguide}
\label{Measurement}



Two sets of Cisco WLAN devices are used, one of which is set as the access point (AP) and the other one is set as the MS. Both of them are set to work at the frequency of 2.412$GHz$, which is also called channel $1$. Due to the effects of coupling loss and transmission loss of leaky waveguide, the signal strength leaked from the slots cannot meet the requirements of CBTC systems if the input power is not high enough.  As a result, the output power of the AP taken as the transmitter is set as 30$dBm$. Through a coupling unit, the signal of the AP is injected into leaky waveguide. The MS taken as the receiver is located on a measurement vehicle with a panel antenna to receive the signals leaked from the leaky waveguide. The gain of the antenna is 11$dBi$ and the beam width is $30 ^{\circ}$.

\begin{figure}[tp]
	\centering
	\subfigure[]{\label{TP}\includegraphics[width=0.28\textwidth]{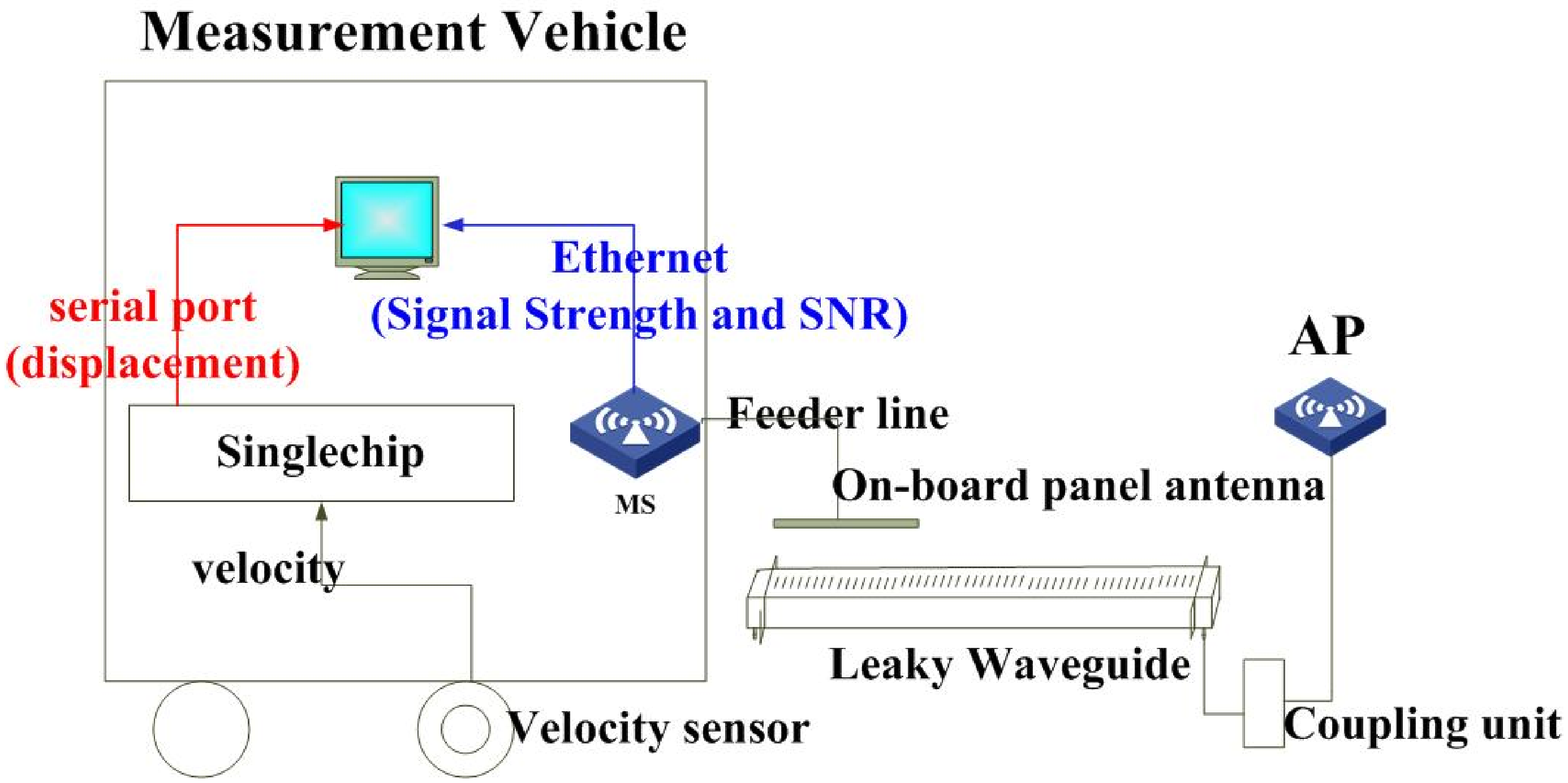}}
	\subfigure[]{\label{TE}\includegraphics[width=0.2\textwidth]{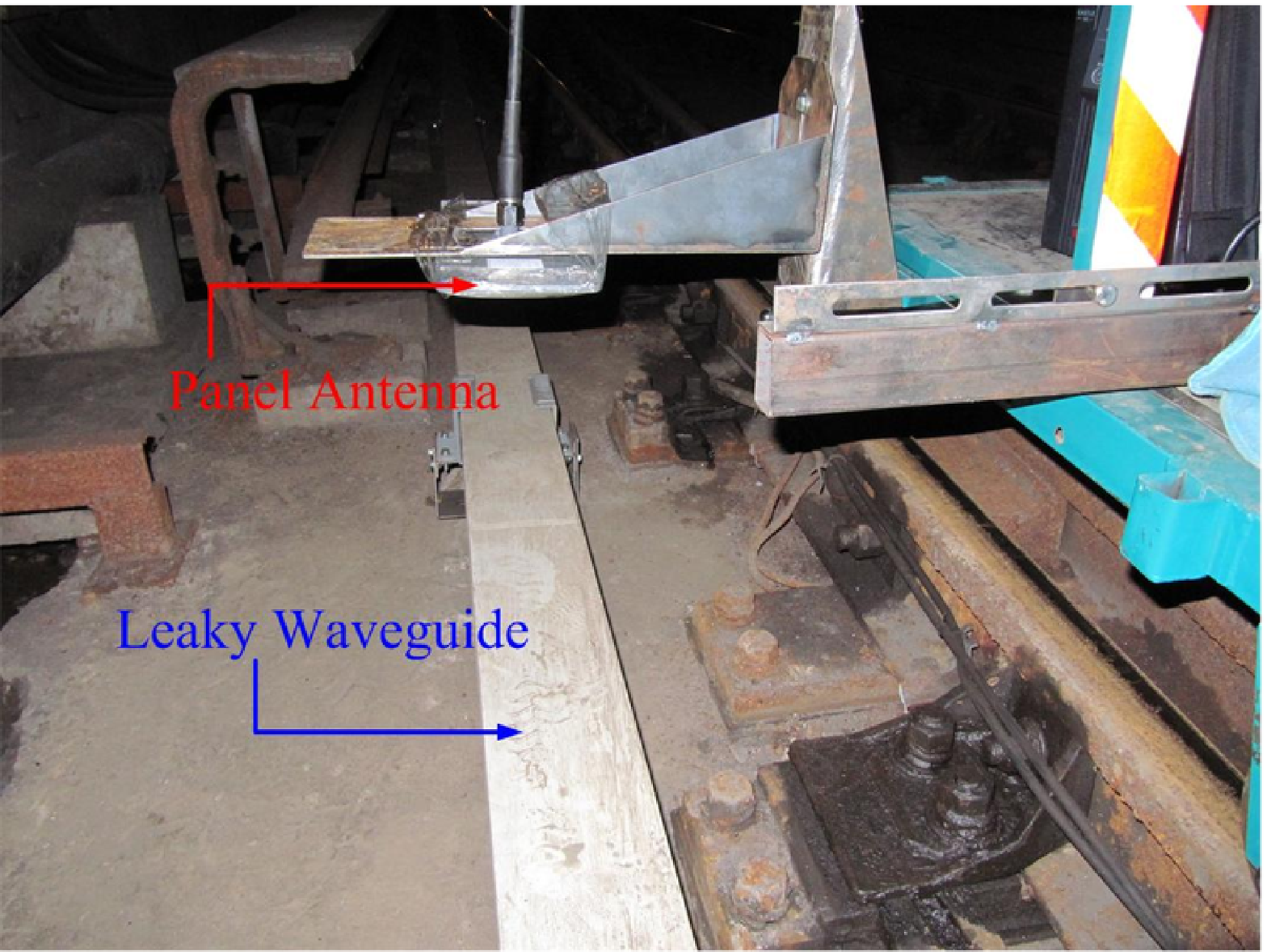}}
	\caption{Measurement campaign. (a) The measurement equipment used in the real field CBTC channel measurements. (b) The measurement scenario.}
	\label{MM}
\end{figure}

Fig. \ref{TP} shows the measurement equipment, and Fig. \ref{TE} shows the measurement equipment used in our real field measurements. The location of the receiver is obtained through a velocity sensor installed on the wheel of the measurement vehicle, which can detect the real-time velocity, and the resolution of position is millimeter per second. In addition, according to the definition of Doppler Frequency shift \cite{Wireless_Communicaitons_Principles_and_Practice} \cite{The_Mobile_Radio_Propagation_Channel} and the characteristics of leaky waveguide, the angle between the direction of motion of the mobile station and the direction of arrival of the wave is almost $90$ degrees, and then the Doppler frequency shift is almost 0, which we can ignore.

The measurements were performed at the section from Rongjingdong Station to Tongjinan Station of Beijing Subway Yizhuang Line. Due to the limits of subway line and the requirements of trains especially the Bogie, the height of the receiving antenna is set as 330$mm$. The length of one section of leaky waveguide is $300m$, which determines the experimental zone in our measurements.

\section{The Finite-State Markov Chain Channel Model}
\label{FSMC_model}

To capture the characteristics of leaky waveguide channels in CBTC systems, we define channel states according to the different received SNR levels, and use an FSMC to track the state variation. In this section, we first describe the FSMC model, followed by the determination of key model parameters, including  SNR levels and SNR distribution.

\subsection{The Finite State Markov Channel Model}
The time axis is divided into slots of equal duration. Let $\gamma_k$ denote the SNR of the received signal in time slot $k$, whose range can be obtained from the experimental data. The range of SNR is partitioned into $N$ non-overlapping levels with thresholds $\{\Gamma_{n}, n = 0, 1, 2, 3, ..., N\}$, where $\Gamma_{0}$ and $\Gamma_{N}$ can be measured. The channel state $\gamma_k=s_{n}$ when the SNR of the received signal belongs to the range $[\Gamma_{n-1}, \Gamma_{n})$. Then the received
SNR can be modeled as a random variable $\gamma$ evolving according to a finite-state Markov chain, and the transition probability $p_{n,j}$ can be shown as follows.
\begin{equation}
p_{n,j}=P_{r}\{\gamma_{k+1}=s_{n} \mid \gamma_{k}=s_{j}\},
\label{STP2}
\end{equation}
where $k = 1, 2, 3, ..., $ and $n, j \in \{1 ,2, ..., N\}$.

According to the property of first-order Markov chain, we assume that each state can only transit to the adjacent states, which means $p_{n,j}=0$, if $\mid{n-j}\mid>1$. With the definition, we can define a $K\times K$ state transition probability matrix $\textbf{P}$ with elements $p_{n,j}$.

Due to the effect caused by the transmission loss of leaky waveguide \cite{Leaky_Waveguide_for_Train-to-Wayside_Communication-Based_Train_Control}, the amplitude of SNR depends on the location of the receiving antenna.  As a result, the transition probability from the high channel state to the low channel state is different when the location of the receiving antenna changes, which means that the Markov state transition probability is related to the location of  the receiver. Therefore, only one state transition probability matrix, which is independent of the location of the receiver, may not accurately model the leaky waveguide channels. Thus, we divide the leaky waveguide into $L$ intervals and one state transition probability matrix is generated for each interval. Specifically, $\textbf{P}^{l}, l \in \{1, 2, ..., L\}$, is the state transition probability matrix corresponding to the $l$th interval, and the relationship between the transition probability and the location of the receiver can be built. Then, the element of $\textbf{P}^{l}$, defined as $p_{n, j}^{l}$, is the state transition probability from state $s_{n}$ to state $s_{j}$ in the $l$th interval. And $p_{n}^{l}$ is defined as the probability of being in state $n$ in the $l$th interval.

Based on the measurement results, we can determine the value of the state probability $p_{n}^{l}$ and the state transition probability $p_{n,j}^{l}$.

\subsection{Determine the SNR Level Thresholds of the FSMC Model}
Determining the thresholds of SNR levels is the key factor that affects the accuracy of the FSMC model. There are many methods to select the SNR level boundaries, among which the equiprobable partition method is frequently used in previous works \cite{Finite_state_Markov_channel_a_useful_model_for_radio_communication_channels,Finite-state_Markov_modeling_of_correlated_Rician-fading_channels,Fast_simulation_of_diversity_Nakagami_fading_channels_using_finite-state_Markov_models}. As non-uniform amplitude partitioning can be useful to obtain more accurate estimates of system performance measures \cite{Finite-state_Markov_modeling_of_fading_channels_a_survey_of_principles_and_applications}, we choose the Lloyd-Max technique \cite{Least_squares_quantization_in_PCM} instead of the equiprobable method to partition the amplitude of SNR in this paper. Lloyd-Max is an optimized quantizer, which can decrease the distortion of scalar quantization.

Firstly, a distortion function $D$ is defined.
\begin{equation}
D=\sum^{N}_{n=1}\int ^{\Gamma_{n}}_{\Gamma_{n-1}}f(\tilde{\Gamma}_{n}-\gamma)p(\gamma)d\gamma,
\label{LLM1}
\end{equation}
where $\tilde{\Gamma}_{n}$ is the quantized value of SNR whose amplitude is in the range $[\Gamma_{n-1} ~~\Gamma_{n})$, $f(\cdot)$ is the error criterion function, and $p(\gamma)$ is the probability distribution function of SNR. The distortion function can be minimized through optimally selecting $\tilde{\Gamma}_{n}$ and ${\Gamma}_{n}$.


The error criterion function $f(x)$ is  often taken as $x^{2}$ \cite{Digital_communications}. And the necessary conditions for minimum distortion are obtained by differentiating $D$ with respect to ${\Gamma_{n}}$ and ${\tilde{\Gamma_{n}}}$ as follows.

\begin{eqnarray}
\label{LLM4}
&\Gamma_{n}=\frac{\tilde{\Gamma}_{n}+\tilde{\Gamma}_{n+1}}{2},\\
\label{LLM5}
&\int^{\Gamma_{n}}_{\Gamma_{n-1}}(\tilde{\Gamma}_{n}-\gamma)p(\gamma)d\gamma=0.
\end{eqnarray}

As mentioned above, we partition the amplitude of SNR into $N$ levels, and there are $N+1$ corresponding thresholds $\{\Gamma_{n}, n= 0, 1, 2, 3, ..., N\}$. Generally, the first and last thresholds are known, which are denoted by the minimum and maximum measurement values of SNR. Furthermore, the Lloyd-Max algorithm is used to divide $2^{r}$ levels, which means $N=2^{r}, r = 1, 2, 3, ....$, and $N$ is an even number. As a result, since $\Gamma_{0}$ and $\Gamma_{N}$ are known, based on \eqref{LLM5}, all elements of $\{\Gamma_{n}\}$ can be obtained.

According to the calculated $\{\Gamma_{n}\}$, combined with \eqref{LLM4} and \eqref{LLM5}, we can update the value of $\{\Gamma_{n}\}$ until the value of $D$ is the minimum, and the optimal thresholds of the SNR levels can be obtained. As $p(\gamma)$ is still unknown, we should discuss the distribution of SNR in the following subsection according to the real field measurement data, which is the last step to obtain the thresholds of SNR levels.

\subsection{Determine the Distribution of SNR}

Deriving the distribution of SNR is the crucial step of partitioning the levels of SNR. In fact, there are some classic models to describe the distribution of signal strength, such as Rice, Rayleigh, Nakagami, Weibull and Log-normal, and then the corresponding models of SNR can also be obtained \cite{Digital_Communication_over_Fading_Channels}. We firstly derive the distribution of the signal strength in order to determine the distribution model of SNR.

The Akaike’s Information Criteria with a correct (AICc) is adopted to get the approximate distribution model of the signal
strength from the five classic models (candidate models) mentioned above which are also used in \cite{A_statistical_model_for_indoor_office_wireless_sensor_channels}. Since our channel model is related to the location of the receiver, the leaky waveguide should be divided into $L$ intervals. And we apply AICc for each candidate model in every interval, which is defined as follows \cite{Model_selection_and_multimodel_inference}.

\begin{equation}
\begin{aligned}
&AIC_{i,j}=-2\sum_{n=1}^{N_{i}}\log_{e}(l(\widehat{\theta}_{i,j}\vert x_{i,n}))+2k_{j},\\
&AICc_{i,j}=AIC_{i,j}+\frac{2k_{j}(k_{j}+1)}{N_{i}-k_{j}-1},
\end{aligned}
\end{equation}
where $i$ means the $i$th interval, $j$ means the $j$th candidate model, $\widehat{\theta}_{i,j}$ means the estimated parameters of the $j$th candidate model for the $i$th interval using the maximum likelihood estimator (MLE), $x_{i,n}$ is the $n$th sample of the $i$th interval, $k_{j}$ is the number of parameters of the $j$th candidate model, $N_{i}$ is the total number of samples of the $i$th interval.

As a result, we can select the most appropriate model based on the frequency of the minimum AICc value of different candidate models. In order to obtain enough data for each interval and ensure the accuracy of the model, we set the length of each interval as $40$ wavelengths of WLANs \cite{A_statistical_model_for_indoor_office_wireless_sensor_channels}. Based on the frequencies of AICc of different distributions in the real field measurements, we observe that the Log-normal distribution provides the best fit compared to other distributions. As a result, we can define $p(\gamma)$ as the Log-normal distribution.

After the distribution of the signal strength is obtained, according to \cite{Digital_Communication_over_Fading_Channels}, we can derive the distribution of SNR.
\begin{equation}
p(\gamma)=\frac{\xi}{\sqrt{2\pi}\sigma \gamma}\left[ -\frac{10\log_{10} \gamma-\mu}{2\sigma^2}\right],
\label{SNRfunc}
\end{equation}
where $\gamma$ is the SNR of the received signal, $\xi=10/ \ln10=4.3429$, $\mu$ and $\sigma$ are the mean and standard deviation of $10\log_{10} \gamma$, respectively. In fact, $\mu$ and $\sigma$ can be calculated when applying AICc through the maximum likelihood estimator for each interval.

\section{Real Field Measurement Results and Discussions}
\label{discussion}
In this section, we compare our FSMC model with real field test results to illustrate the accuracy  of the model. The effects of distance interval in the proposed model are discussed. The number of states in our model is set as $4$. In order to obtain the effects of distance intervals on the model, we choose the intervals as $5m$, $10m$, $20m$, $25m$, $40m$, $50m$, $100m$ and $300m$. We perform measurements in the viaduct section of Beijing Subway Yizhuang Line for dozens of times so that enough data can be captured. The accuracy of the FSMC model is verified through another set of measurement data.

Based on the measurement data, (\ref{LLM4}), (\ref{LLM5}) and (\ref{SNRfunc}), we derive the thresholds $\{\Gamma_{n},n=0,1,2,...,N\}$ of SNR in each distance interval. For example, if the  distance interval is $5m$, at the location $[295m, 300m]$, the thresholds are $[35, 37.3494, 38.7291, 40.0784, 42]$;
 When the distance interval is $300m$, at the location $[0m, 300m]$, the thresholds are $[28,41.0254,43.6633,46.0723,49]$. As the distance intervals are different, the range of SNR is different and it brings different thresholds, which can provide more accurate model.

\begin{table}[tp]
\begin{center}
\caption{The state transition probabilities of the FSMC model and the measurement data with 4 state and 5m interval at the location $(15m-20m)$}
\label{STPM}
\begin{tabular}{|c|c|c|c|c|c|c|}
\hline
   &\multicolumn{3}{c}{The FSMC Model}\vline&\multicolumn{3}{c}{The Measurement Data}\vline\\ \cline{2-7}
   &$p_{k,k-1}$&$p_{k,k}$&$p_{k,k+1}$&$p_{k,k-1}$&$p_{k,k}$&$p_{k,k+1}$\\ \hline
k=1&-&0.736&0.263&-&0.75&0.25\\\hline
k=2&0.253&0.503&0.243&0.25&0.75&0.25\\\hline
k=3&0.210&0.587&0.202&0.2&0.6&0.2\\\hline
k=4&1&0&-&1&0&-\\\hline

\end{tabular}
\end{center}
\end{table}

\begin{figure}
  \centering
  \subfigure[]{\includegraphics[width=0.24\textwidth]{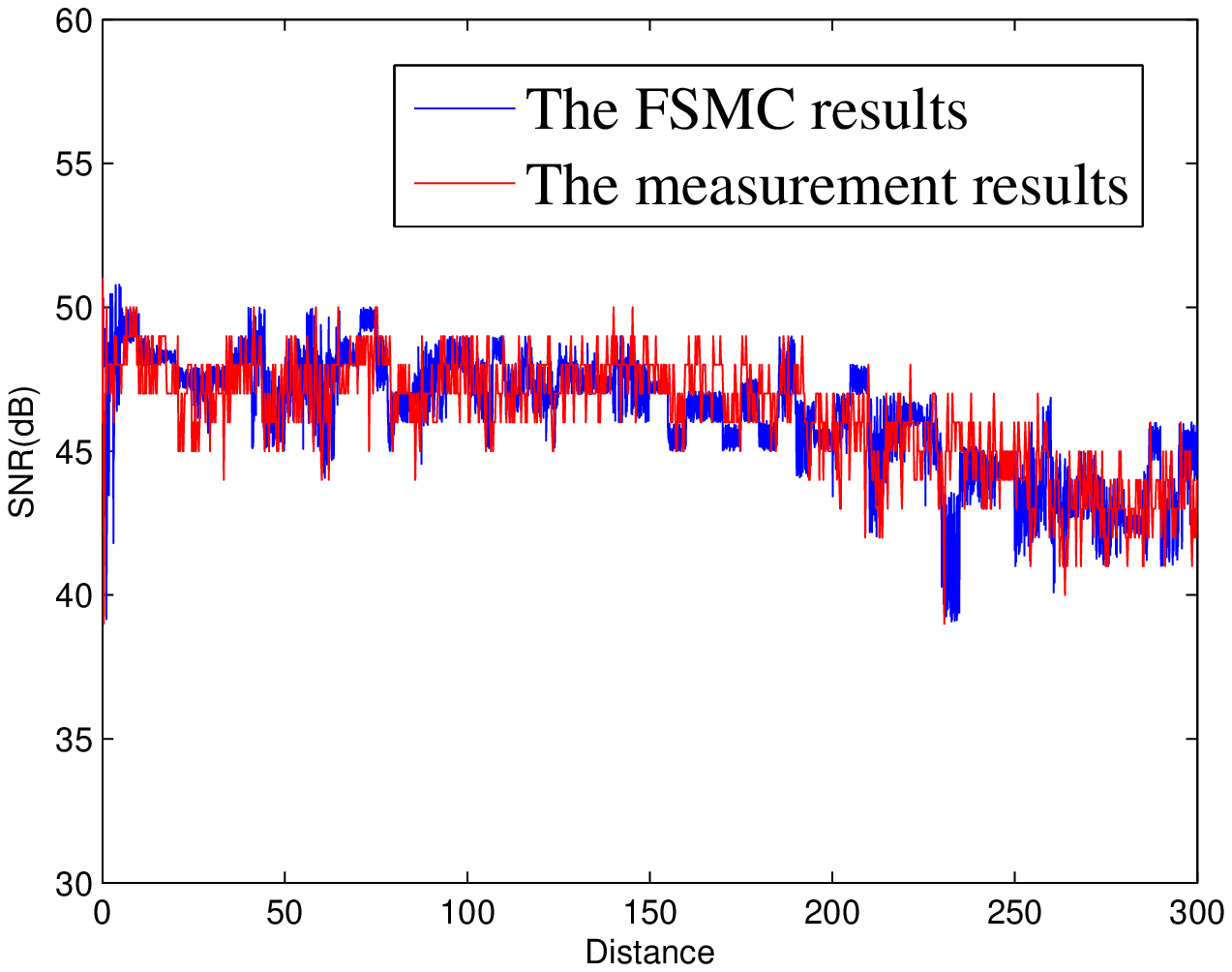}}
  \subfigure[]{\includegraphics[width=0.24\textwidth]{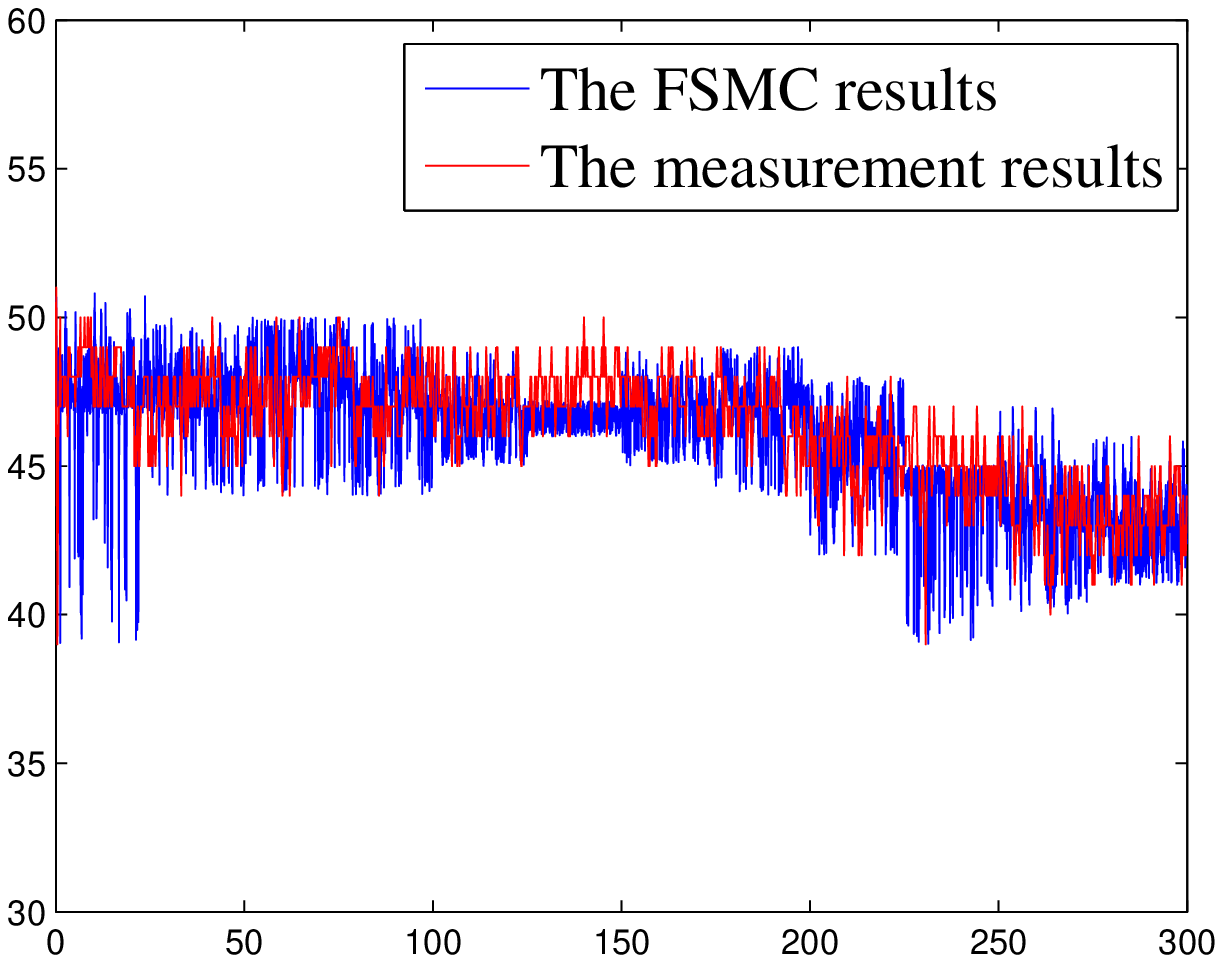}}
  \caption{Simulation results generated from the FSMC model and experimental results from real field measurements. (a) The distance interval is $5m$ (b) The distance interval is $50m$.}
  \label{simulation_results}
  \includegraphics[width=0.33\textwidth]{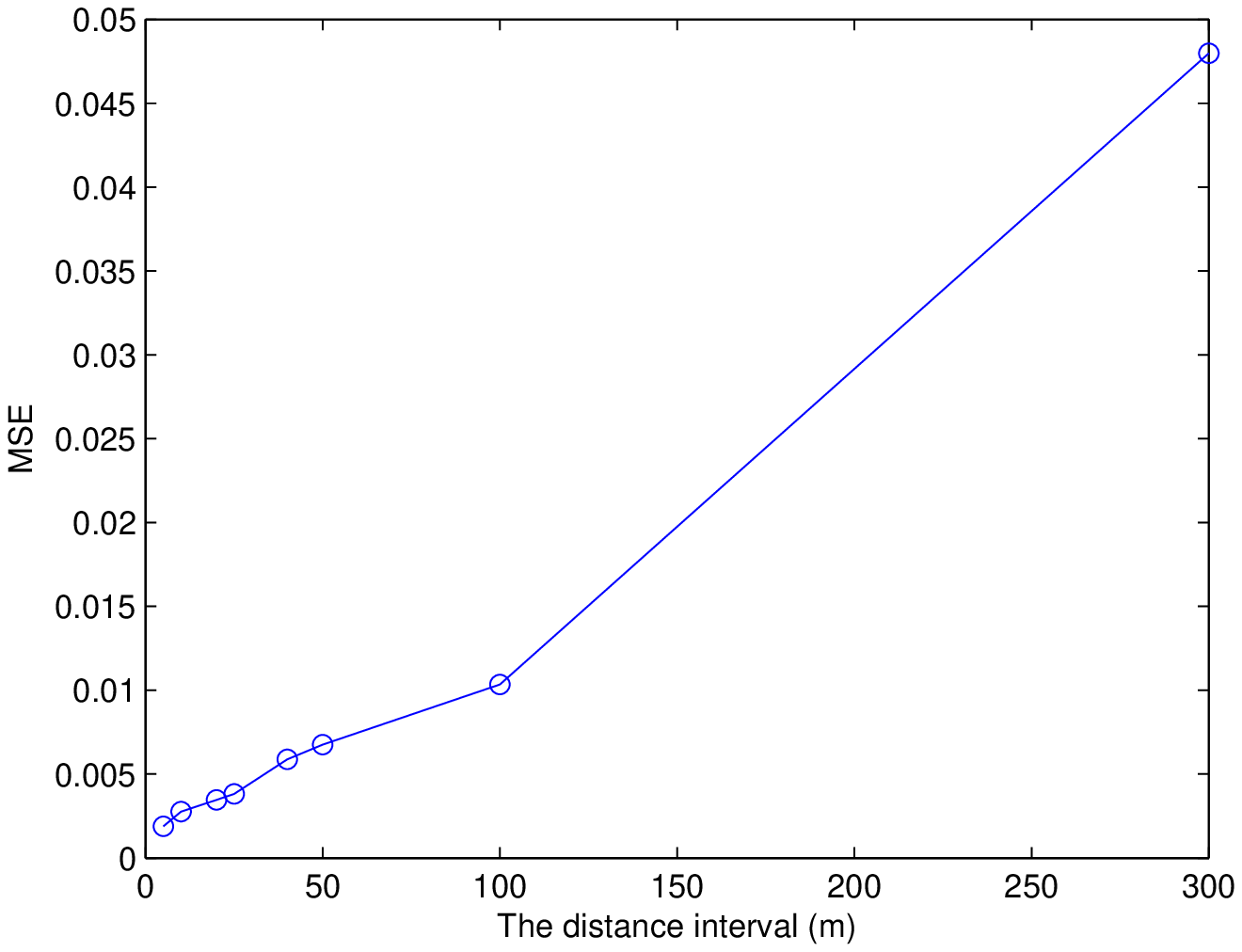}
\caption{The mean square error (MSE) between the FSMC model and the experimental data with $4$ states.}
\label{MSE}
\end{figure}

After we get the thresholds,  we can get the state probabilities and the state transition probabilities from the real field data. Table \ref{STPM} illustrates the state transition probabilities of the FSMC model and the measurement data at the same location $(15m-20m)$, when there are four states and the distance interval is $5m$. Fig. \ref{simulation_results} shows the simulation results generated from our FSMC model and the experimental results from real field measurements. We can observe there is more agreement between them when the distance interval is $5m$ than that with $50m$ distance interval. Next, we derive the mean square error (MSE) to measure the degrees of approximation, shown in Fig. \ref{MSE}. As we can see from Fig. \ref{MSE}, when the distance interval increases, the MSE also increases, which means that the accuracy of the model decreases. However, MSE is almost the same when the distance interval is $5m, 10m, 20m$ and $25m$. The reason is that, as the signal leaked from the slots of leaky waveguide is stable, the range of SNR does not change significantly when the distance interval increases to a limited extent.  From the results of Fig. \ref{MSE}, we can see that the FSMC model with 4 states and $25m$ distance interval can provide an accurate enough channel model for leaky waveguide channels in CBTC systems.

\section{Conclusions and Future work}
\label{conclusion}

We have proposed an FSMC model for leaky waveguide of CBTC systems. The proposed FSMC channel model takes train locations into account to have a more accurate channel model. The overall leaky waveguide is divided into intervals, and an FSMC model is designed in each interval. The accuracy of the proposed model has been illustrated by the simulation results generated from the proposed model and the real field measurements. In addition,  we have shown that the distance interval have impacts on the accuracy of the proposed FSMC model. Future work is in progress to study the effects of  wireless channels on the control performance of CBTC systems based on the proposed channel model.

\section*{Acknowledgement}

This paper was supported by grants from the National Natural Science Foundation of China (No.61132003), the National High Technology Research and Development Program of China (863 Program) (2011AA110502), and projects (No. RCS2011ZZ007, RCS2012ZQ002, 2013JBM124, 2011JBZ014, RCS2010ZZ003, RCS2012K010).

\bibliographystyle{IEEEtran}
\bibliography{ref}

\begin{thebibliography}{10}

\bibitem{What_is_communication-based_train_control}
R.~Pascoe and T.~Eichorn, ``What is communication-based train control?,'' {\em
  IEEE Veh. Tech. Mag.}, vol.~4, pp.~16 --21, Dec. 2009.

\bibitem{CBTC_Standard}
IEEE, ``Standard for communications-based train control ({CBTC}) performance
  and functional requirements,'' {\em IEEE Std 1474.1-2004 (Revision of IEEE
  Std 1474.1-1999)}, pp.~0\_1 --45, 2004.

\bibitem{Leaky_Waveguide_for_Train-to-Wayside_Communication-Based_Train_Control}
M.~Heddebaut, ``Leaky waveguide for train-to-wayside communication-based train
  control,'' {\em IEEE Trans.\ Veh.\ Tech.}, vol.~58, pp.~1068 --1076, Mar.
  2009.

\bibitem{ZHULI_FSMC}
L.~Zhu, F.~Yu, B.~Ning, and T.~Tang, ``Cross-layer handoff design in
  {MIMO}-enabled {WLANs} for communication-based train control ({CBTC})
  systems,'' {\em IEEE J.\ Sel.\ Areas\ Commun.}, vol.~30, pp.~719 --728, May
  2012.

\bibitem{ZFBT12}
L.~Zhu, F.~R. Yu, B.~Ning, and T.~Tang, ``{Handoff Performance Improvements in
  MIMO-Enabled Communication-Based Train Control Systems},'' {\em {IEEE
  TRANSACTIONS ON Intelligent Transportation Systems}}, vol.~13, no.~2,
  pp.~582--593, 2012.

\bibitem{Characterization_of_Large_Scale_Fading_for_the_2_4_GHz_Channel_in_Obstacle_Dense_Indoor_Propagation_Topologies}
T.~Chrysikos and S.~Kotsopoulos, ``Characterization of large-scale fading for
  the 2.4 ghz channel in obstacle-dense indoor propagation topologies,'' in
  {\em Proc. IEEE VTC'12 fall}, pp.~1--5, 2012.

\bibitem{5_GHz_Band_Vehicle_to_Vehicle_Channels_Models_for_Multiple_Values_of_Channel_Bandwidth}
Q.~Wu, D.~Matolak, and I.~Sen, ``5-ghz-band vehicle-to-vehicle channels: Models
  for multiple values of channel bandwidth,'' {\em IEEE Trans.\ Veh.\ Tech.},
  vol.~59, no.~5, pp.~2620--2625, 2010.

\bibitem{Channel_Modeling_for_Vehicle_To_Vehicle_Communications}
D.~Matolak, ``Channel modeling for vehicle-to-vehicle communications,'' {\em
  IEEE Comm. Mag.}, vol.~46, no.~5, pp.~76--83, 2008.

\bibitem{Finite_state_Markov_channel_a_useful_model_for_radio_communication_channels}
H.~S. Wang and N.~Moayeri, ``Finite-state {Markov} channel-a useful model for
  radio communication channels,'' {\em IEEE Trans.\ Veh.\ Tech.}, vol.~44,
  pp.~163 --171, Feb. 1995.

\bibitem{Finite-state_Markov_modeling_of_correlated_Rician-fading_channels}
C.~Pimentel, T.~Falk, and L.~Lisboa, ``Finite-state {Markov} modeling of
  correlated {Rician}-fading channels,'' {\em IEEE Trans.\ Veh.\ Tech.},
  vol.~53, pp.~1491 -- 1501, Sept. 2004.

\bibitem{Fast_simulation_of_diversity_Nakagami_fading_channels_using_finite-state_Markov_models}
C.~Iskander and P.~Mathiopoulos, ``Fast simulation of diversity {Nakagami}
  fading channels using finite-state {Markov} models,'' {\em IEEE Trans.\
  Broadcasting.}, vol.~49, pp.~269 -- 277, Sept. 2003.

\bibitem{Wireless_Communicaitons_Principles_and_Practice}
T.~S.Rappaport, {\em Wireless Communications : Principles and Practice}.
\newblock Upper Saddle River, NJ: Prentice Hall, 1999.

\bibitem{The_Mobile_Radio_Propagation_Channel}
J.~D. Parsons, {\em The Mobile Radio Propagation Channel}.
\newblock Hoboken, NJ: Wiley Interscience, 2000.

\bibitem{Finite-state_Markov_modeling_of_fading_channels_a_survey_of_principles_and_applications}
P.~Sadeghi, R.~Kennedy, P.~Rapajic, and R.~Shams, ``Finite-state {Markov}
  modeling of fading channels - a survey of principles and applications,'' {\em
  IEEE Signal Proc. Mag.}, vol.~25, pp.~57 --80, Sep. 2008.

\bibitem{Least_squares_quantization_in_PCM}
S.~Lloyd, ``Least squares quantization in {PCM},'' {\em IEEE Trans.\ Inform.\
  Theory}, vol.~28, pp.~129 -- 137, Mar. 1982.

\bibitem{Digital_communications}
J.~Proakis, {\em Digital communications}.
\newblock McGraw-Hill, 1995.

\bibitem{Digital_Communication_over_Fading_Channels}
M.-S.~A. Marvin K.~Simon, {\em Digital Communication over Fading Channels}.
\newblock John Wiley \& Sons, 2005.

\bibitem{A_statistical_model_for_indoor_office_wireless_sensor_channels}
S.~Wyne, A.~Singh, F.~Tufvesson, and A.~Molisch, ``A statistical model for
  indoor office wireless sensor channels,'' {\em IEEE Trans.\ Wireless
  Commun.}, vol.~8, pp.~4154 --4164, Aug. 2009.

\bibitem{Model_selection_and_multimodel_inference}
A.~D. Burnham~KP, {\em Model selection and multimodel inference: a practical
  information-theoretic approach}.
\newblock Springer, 2002.

\end{thebibliography}

\end{document}